Competing quantum Hall phases in the second Landau level in the low-density limit

W. Pan[1], A. Serafin[2], J.S. Xia[2], L. Yin[2], N.S. Sullivan[2], K.W. Baldwin[3], K.W. West[3], L.N. Pfeiffer[3], and D.C. Tsui[3]

[1] Sandia National Laboratories, Albuquerque, New Mexico, USA 87185
[2] University of Florida and National High Magnetic Field Lab, Gainesville, FL 32611
[3] Princeton University, Princeton, New Jersey, USA 08544

Abstract:

In this Rapid Communication we present the results from two high-quality, low-density GaAs quantum wells. In sample A of electron density n=5.0×10$^{10}$ cm$^{-2}$, anisotropic electronic transport behavior was observed at ν=7/2 in the second Landau level. We believe that the anisotropy is due to the large Landau level mixing effect in this sample. In sample B of density 4.1×10$^{10}$ cm$^{-2}$, strong 8/3, 5/2, and 7/3 fractional quantum Hall states were observed. Furthermore, our energy gap data suggest that, similar to the 8/3 state, the 5/2 state may also be spin unpolarized in the low-density limit. The results from both samples show that the strong electron-electron interactions and a large Landau level mixing effect play an import role in the competing ground states in the second Landau level.

When a high-quality two-dimensional electron system (2DES) is placed in high magnetic fields and cooled to very low temperature, many novel quantum ground states appear, for example, the integer and fractional quantum Hall effect states [1], the composite fermions (CF) Fermi sea state [2,3], the pinned Wigner crystal states [4,5], the stripe and bubble states [6-9], etc. The fundamental physics underlying these quantum phenomena and the competition among their ground states is the strong electron-electron interactions.

In recent years, exotic electron physics in the second (N=1) Landau level has been the center of solid state physics research, due to the exciting proposal of using non-Abelian fractional quantum Hall effect (FQHE) states in fault-resistant topological quantum computation [10]. Unlike in the first or higher Landau levels, it is in this second Landau level where the conventional models, such as the weakly interacting CF model, fail to explain the origin of the FQHE at the Landau level filling ν=5/2 [11,12]. Instead, a pairing mechanism of CFs [13], not unlike the Cooper-pairs of electrons in a superconductor, has to be invoked to explain the 5/2 FQHE state, and its particle-hole conjugate state at ν=7/2. Under this pairing picture, it is generally believed that the ground states at ν=5/2 and 7/2 are spin polarized and isotropic. Late, in a series of experimental and theoretical work [14-16], it was shown that the paired FQHE states are very close to the boundary of the stripe phase. A small in-plane magnetic field can drive the state from the FQHE state to an anisotropic state.

To date, studies of the FQHE states in the second Landau level have mainly been carried out in the high electron density regime, where the electron mobility is the highest. Only recently, with the advance of high-quality, low-density MBE growth, experiments have been pushed to the low-density regime [17-19], where the electron-electron interactions are strong and the Landau level mixing parameter, defined by $\kappa = e^2/\varepsilon l_B/\hbar\omega_c$, is large. Here, $l_B = (\hbar e/B)^{1/2}$ is the magnetic length and $\omega_c = eB/m$ the cyclotron frequency. All other parameters have their normal meanings. It has been shown that a large Landau level mixing effect strongly affects the electron physics in the second Landau level [20-24].

In this Rapid Communication, we present our recent results obtained in very low-density and high-quality 2DES realized in symmetrically doped GaAs quantum wells. Our result shows that the 7/2 state, a FQHE state in high-density samples, becomes anisotropic in a sample of density n = $5.0\times10^{10}$ cm$^{-2}$. In another sample with a lower electron density of $4.1\times10^{10}$ cm$^{-2}$, strong 8/3, 5/2 and 7/3 FQHE states were observed. Comparison with previous data suggests that the 5/2 state may be spin-unpolarized in this sample. Our results demonstrate that in the low-density regime the strong electron-electron interactions and large Landau level mixing effect play an important role in competing ground states in the second Landau level.

The samples we examined are symmetrically modulation-doped GaAs quantum wells. In sample A, the low temperature electron density and mobility are n=$5\times10^{10}$ cm$^{-2}$ and µ = $10\times10^6$ cm$^2$/Vs, respectively. The quantum well width is 60nm. In sample B, n=$4.1\times10^{10}$ cm$^{-2}$ and µ = $9\times10^6$ cm$^2$/Vs. The well width is 65nm. All measurements were carried out in a dilution refrigerator with the lowest base temperature of ~ 12 mK. A low frequency (~ 7Hz) lock-in technique was utilized to measure the magnetoresistance $R_{xx}$ and $R_{yy}$. The excitation current was normally 2 or 5 nA.

Figure 1 shows the $R_{xx}$ and $R_{yy}$ traces taken in sample A at T = 110 mK. At this high temperature, electronic transport is isotropic in the whole magnetic (B) field range. The minor difference between $R_{xx}$ and $R_{yy}$ is due to the different contact combinations we used to measure $R_{xx}$ and $R_{yy}$. As the sample temperature is cooled to 16mK, as shown in Fig. 2, unlike in high density samples where the half-filled states at ν = 9/2, 11/2 … become strongly anisotropic [8,9], here in this low-density sample, they remain more or less isotropic. This isotropic phase at these half Landau level fillings is probably due to a lower melting temperature of the stripe phase in the low electron density limit [25]. It is possible that at yet lower temperatures, anisotropy will develop at 9/2, 11/2, etc. In the second Landau level, the 7/2 state, which is a fractional quantum Hall state [26] and isotropic in high density samples, now becomes noticeably anisotropic at T ~ 16mK, as shown in Fig. 2. The ratio of $R_{xx}/R_{yy}$ is about 2. We emphasize here that this anisotropic 7/2 state was observed with the sample carefully placed perpendicular to the external magnetic field and there was no intentionally applied in-plane B field. The anisotropic transport behavior at 7/2 is very sensitive to the sample temperature. By a mere increase of 8 mK, from ~ 16 mK to ~ 24 mK, the 7/2 state becomes fully isotropic. Around 5/2, $R_{xx}$ and $R_{yy}$ also differ from each other. This discrepancy, however, probably is due to an extrinsic cause. In fact, the ratio of $R_{xx}/R_{yy}$ at 5/2 shows very weak temperature dependence, decreasing from ~ 1.5 at 16 mK to only ~ 1.4 at 110 mK, in contrast to that at ν=7/2. In this regard, we believe that the 5/2 state remains isotropic in this sample.

Here we observed an anisotropic electronic transport behavior at ν=7/2 in a high-quality dilute two-dimensional *electron* system. In two-dimensional *hole* systems, anisotropy at ν=7/2 has been observed before [27-29]. There, it was believed that the strong electron-electron interactions and/or spin-orbit coupling were responsible for the observed anisotropy. We believe that this strong electron-electron interaction origin and the interaction-induced large Landau level mixing effect are also responsible for the observed anisotropy in our dilute electron system. In the high-density limit, the FQHE at ν=7/2 is due to pairing of composite fermions [13]. This pairing mechanism is made possible by the flux attachment to electrons. In so doing, the effective interaction between CFs becomes attractive [30]. In the low-density limit, the flux attachment may not be able to fully compensate the strong repulsive interactions between the electrons. As a result, the effective interaction between two CFs remains repulsive. Moreover, a large κ can bring the lowest unoccupied N=2 Landau level energetically closer to the N=1 Landau level, and give electron correlations some of the character of the N=2 Landau level [31]. This contribution from the N=2 Landau level wavefunction, together with the repulsive residual interactions between the CFs, can be expected to turn the ν=7/2 state anisotropic. In this regard, we note that it was demonstrated in a recent theoretical paper that there existed instability to stripe ordering in the presence of a large Landau level mixing effect [32]. In particular, the stripe phase prevails when there are two states close to the Fermi energy and when the filling of one or both states is very close to the half-filling. Furthermore, numerical calculations [21] showed that the overlap between a specific exact ground state and a Pfaffian state (which is expected to be also responsible for the 7/2 FQHE state in high-density samples) decreased when κ > 2, suggesting an instability of the Pfaffian state at low densities. In sample A κ = 3.3 at ν=7/2, much larger than κ = 1.3 in the high-density samples of $n=3\times10^{11}$ cm$^{-2}$, where an isotropic ν=7/2 FQHE state has been observed. The authors in Ref. [21] did not calculate the overlap between the exact ground state and an anisotropic state. It remains interesting to observe whether under their numerical model the anisotropic state actually becomes more stable in the large κ regime.

Other mechanisms, for example, the finite thickness effect in the 2DES, may also contribute to the observed anisotropy. On the other hand, although the quantum well width (60nm) in sample A is significantly wider than that (30nm) in high-density samples, the ratio of the quantum well width over the magnetic length at $\nu=7/2$ is almost the same in both types of samples. If anything, this finite quantum well width induced Landau level mixing effect should be of the same order.

Finally, it is known that the short-range Coulomb interactions are softened in 2DES with large κ [33,34]. Consequently, FQHE becomes less stable. Indeed, it has been shown that strong electron-electron interactions can induce a transition from the FQHE liquid state to the pinned Wigner crystal state (a charge-density-wave state) [4,5] in the lowest Landau level in high magnetic fields. It is possible a similar transition can also be induced in the second Landau level. Here, due to the additional node in the Landau level wave function the stripe state (also a charge-density-wave state) may become more stable at $\nu=7/2$ than other charge-density-wave states, such as Winger crystal or bubble states.

Having demonstrated that the strong electron-electron interactions can alter the ground state at $\nu=7/2$ in a dilute two-dimensional electron sample, in the following, we will present our results on the FQHE in the second Landau level in sample B with an even lower density of $4.1\times10^{10}$ cm$^{-2}$.

In Figure 3a, we show the $R_{xx}$ trace between $\nu=2$ and 3, taken at ~12 mK. Unfortunately, we cannot measure $R_{yy}$ in sample B due to a limited number of good Ohmic contacts. Nevertheless, evidence of anisotropy at $\nu=7/2$, measured in non-ideal contact configurations, was observed. What is really surprising in this ultra-low-density sample is that the 8/3, 5/2, and 7/3 states show very strong minima. Similar to previous observation in sample A [17], the 8/3 minimum assumes a lower resistance value compared to that at 5/2 and 7/3. This is consistent with the 8/3 state being spin unpolarized at low electron densities. Besides these three strong FQHE states, developing quantum Hall states are also seen at $\nu=2+1/5$ and $2+4/5$. These features attest the high quality of this sample.

The temperature dependence of $R_{xx}$ was also measured to deduce the energy gap at 5/2. Figure 3b shows the standard Arrhenius plot. From the slope of linear fit, we obtained $\Delta_{5/2} = 54$ mK. The uncertainty in the data fitting, for this sample and other samples we measured, is less than 5%. This is much less than the change in the 5/2 energy gap from 54mK at $4.1\times10^{10}$ cm$^{-2}$ to 15mK at $5.0\times10^{10}$ cm$^{-2}$. In Figure 4, we plot the 5/2 gap as a function of densities, including our previous results, and results from Ref. [19], Ref. [35], and Ref. [36]. The two density points at n ~ $4.4 \times 10^{10}$ cm$^{-2}$ were obtained in Bay 3 at the high B/T facilities of National High Magnetic Field Laboratory in a specimen cut from the same wafer as sample B. Different cooling histories are responsible for the slight difference in the electron density. It is seen that the energy gap first decreases from $4.1\times10^{10}$ to $5.0\times10^{10}$ cm$^{-2}$ and then increases again. This density dependence of energy gap was also observed at ν=8/3 in the second Landau level [17] and at 2/3 and 4/3 in the lowest Landau level, and was generally believed to be due to a spin transition. Following the same argument, the results in Figure 4 suggest that the 5/2 state may not be spin polarized in sample B.

We point out that a spin-polarized 5/2 FQHE state in the high-density regime has been confirmed in two resistively detected NMR experiments [37,38]. On the other hand, the assumption of full spin polarization at ν=5/2 is not always justified [20]. In fact, indication of a possible spin unpolairzed 5/2 state was observed in optical measurements [39,40] and was also suggested in Ref. [41]. Moreover, our experiment was carried out in an unexplored ultra-low-density limit. At the present time, the ultimate origin of a spin-unpolarized 5/2 state is not known. Nevertheless, the large Landau level mixing effect (κ ~ 3.0 at ν=5/2) has to play an important role. In this regard, we note here that it was shown in an early report that a large Landau level mixing effect favored the type of pairing correlations presented in the original spin-singlet hollow-core-model wave function for the 5/2 state [42].

In summary, in a high-quality and low-density GaAs quantum well sample, we observed an anisotropic electronic transport behavior at ν=7/2 in the second Landau level. We believe this anisotropy is due to a large Landau level mixing effect in our sample. In another sample with an even lower electron density, our results seem to suggest that the 5/2 FQHE state may be spin-unpolarized. More studies are needed to understand whether this apparent spin unpolarized ground state is also induced by a large Landau level mixing effect.

We would like to thank W. Kang and N.E. Bonesteel for very helpful discussions. This work was supported by the U.S. Department of Energy, Office of Science, Basic Energy Sciences, Materials Sciences and Engineering Division. Sandia National Laboratories is a multi-program laboratory managed and operated by Sandia Corporation, a wholly owned subsidiary of Lockheed Martin Corporation, for the U.S. Department of Energy's National Nuclear Security Administration under Contract No. DE-AC04-94AL85000. A portion of measurements was carried out at the high B/T facility of the National High Magnetic Field Laboratory. The work at Princeton was supported by the DOE under Grant No. DE-FG02-98ER45683, and partially funded by the Gordon and Betty Moore Foundation as well as the National Science Foundation MRSEC Program through the Princeton Center for Complex Materials (DMR-0819860).

Figure captions:

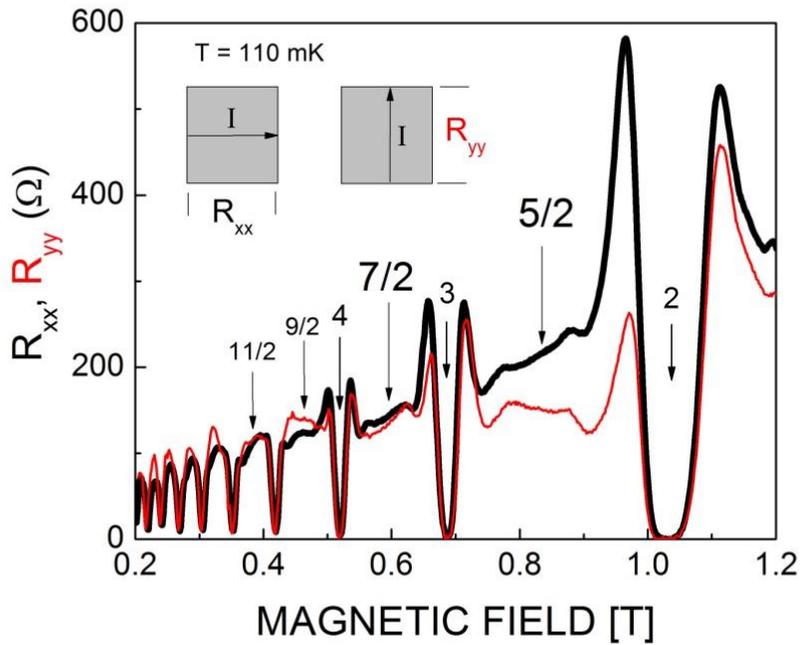

Figure 1: (Color online) Magneto-resistance $R_{xx}$ [thick (black)] and $R_{yy}$ [thin (red)] measured at 110mK in sample A. The arrows mark the positions of integer quantum Hall states at the Landau level fillings ν=2,3,4 and the half-filled states at ν=5/2, 7/2, 9/2, and 11/2. At this high temperature, the electronic transport is isotropic at ν=7/2 and other filling factors -- Rxx and Ryy nearly overlap with each other. The insets show the contact configurations for the $R_{xx}$ and $R_{yy}$ measurements.

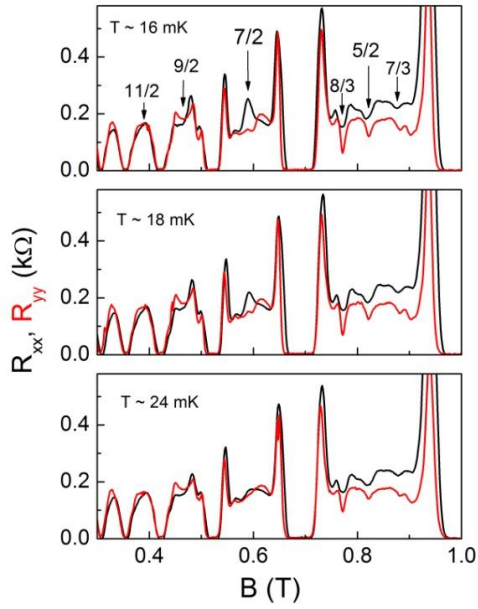

Figure 2: (Color online) $R_{xx}$ and $R_{yy}$ at three temperatures of T ~ 16, 18, and 24mK in sample A. Noticeably anisotropic transport is clearly seen at $\nu=7/2$ at T ~ 16mK. This anisotropy disappears at T ~ 24 mK.

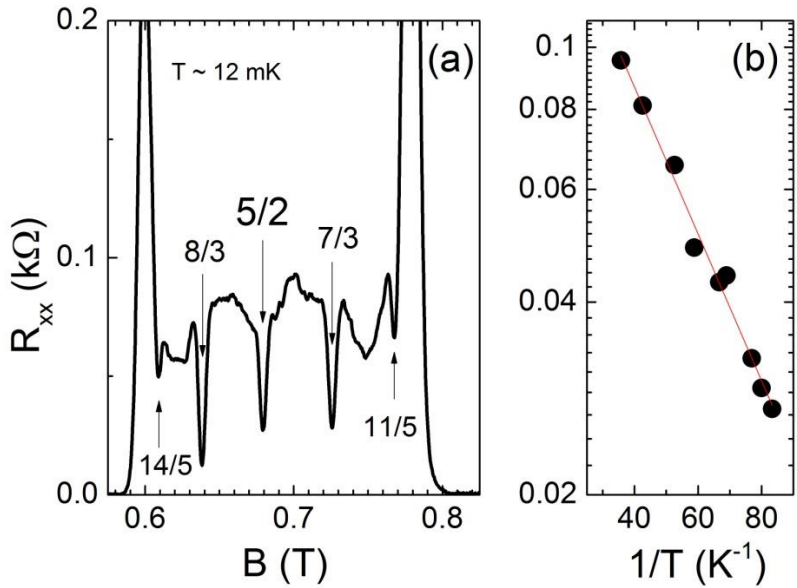

Figure 3: (a) $R_{xx}$ in sample B taken at T ~ 12 mK. The arrows mark the positions of the fractional quantum Hall states at Landau level fillings $\nu=14/5, 8/3, 5/2, 7/3$, and $11/5$. (b) Arrhenius plot for the Rxx minimum at $\nu=5/2$. The line is a linear fit.

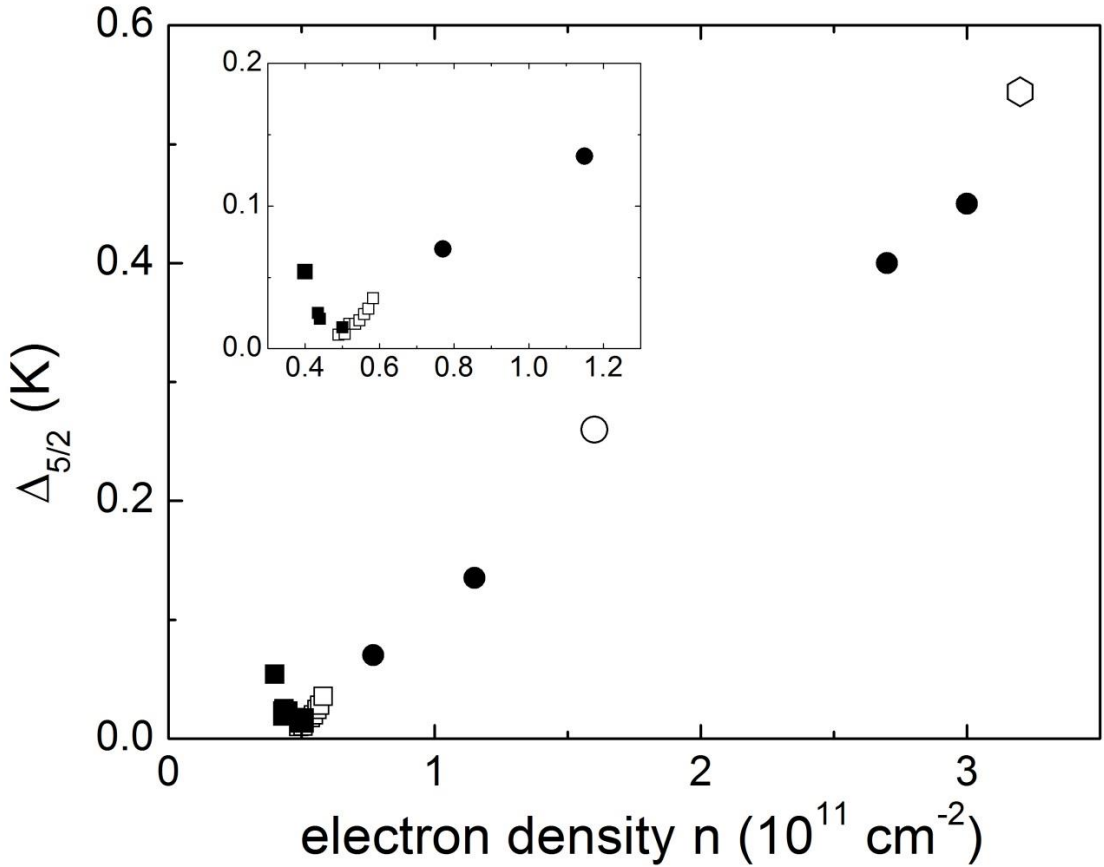

Figure 4: The 5/2 energy gap as a function of density. The four low-density data points (solid squares) were obtained in this experiment. The higher-density points (solid circles) were obtained in the samples from Ref. [17]. Data from Ref. [19] (open squares), Ref. [35] (open hexagon) and Ref. [36] (open circle) are also included. The inset highlights the data points in the low-density regime. The 5/2 energy gap first decreases with increasing density, reaches a minimal value at n ~ $5.0\times10^{10}$ cm$^{-2}$, and then increases with increasing density. This density dependence is consistent with a spin transition in the fractional quantum Hall effect.